\documentclass[12pt, nofootinbib,preprint,showpacs]{revtex4}

\usepackage{amsmath,amssymb}
\usepackage{slashed,mathrsfs}
\usepackage{feynmp,subfigure}
\usepackage{verbatim}
\usepackage{graphicx}
\usepackage{pstricks,pst-coil,pst-node}
\usepackage[sub,ovp]{psfragx}
\usepackage{color}

\newcommand\vev[1]{\langle #1\rangle}

\def\pp{\vskip\baselineskip\noindent}
\def\tr{\textrm{tr}}
\preprint{NSF-KITP-12-016}

\begin{document}
\setlength{\unitlength}{1mm}

\title{Open string axions and the flavor problem}
\author{David Berenstein$^{\dagger, \ddagger}$ and Erik Perkins$^\dagger$ \\
$^\dagger$ {\em Department of Physics, University of California at Santa Barbara, CA 93106}\\
$^\ddagger$ {\em Kavli Institute for Theoretical Physics, Santa Barbara, CA 93106 }}

\begin{abstract}
We consider  extensions of the standard model inspired by intersecting D-brane constructions, in order to address flavor mass textures. We include additional anomalous gauge symmetries, and scalar fields to break them and to generate Froggatt-Nielsen mass terms. Green-Schwarz axions are included to cancel mixed anomalies rendering the models consistent.  At low energies, a residual anomalous global symmetry remains, and its associated pseudo-Goldstone mode becomes the physical axion, which can be interpreted as an axion arising from open string modes. General considerations show that such axions are very common in D-brane models and can be completely incompatible with current bounds.  Astrophysical constraints are placed on the axion both by including neutrino masses in the Froggatt-Nielsen scheme and considering QCD instanton contributions to the axion mass.  We  find  simple models where the axion decay constant is in the allowed range, but only one such minimal model with this property is free from excessive fine tunings elsewhere. We also note that generically addressing flavor textures for the CKM matrix leads to deconstructed extra dimensions.
\end{abstract}


\maketitle

\section{Introduction}
\label{S:Introduction}

The standard model includes no fundamental explanation for the hierarchy of elementary particle masses. The masses, and the Yukawa couplings which determine them, are spread across five orders of magnitude.  This is the problem of flavor in the standard model of particle physics. Here we also include the fact that the CKM matrix is an input to the standard model and is not the result of a calculation. This problem of flavor begs for an explanation from a more fundamental theory of physics. A natural candidate for such a theory is string theory, which in principle would also unify the  gauge interactions with gravity.
As the standard model is weakly coupled at high energies in current accelerator data, it makes sense to take a weakly coupled string theory model to describe the standard model physics plus its extension to flavor physics. For the purposes of this paper we will work in models where the standard model is realized on stacks of D-branes and instead of a top-down approach, we will assume that everything can be described in terms of effective field theory of open string states of spins zero, one half and one, plus massless closed string states as required for anomaly cancelation. All these states are effectively massless at the string scale: we expect all masses to be generated at low energies by  field theory mechanisms.

A natural way to explain a pattern, such as the flavor problem, is in terms of symmetry and a symmetry breaking order parameter. This would explain the hierarchies by having different operators in the standard model transform under different representations of the symmetry group. If the symmetry is continuous, such a symmetry breaking would lead to a massless Goldstone boson as a low energy remnant. Such a possible remnant can be an axion and it is interesting to find models where the axion is related to this symmetry breaking of a flavor symmetry. It has been argued that within string theory and perhaps in any theory of gravity, there can be no global symmetries ( see
\cite{Banks:2010zn} for a modern analysis of this argument), so any symmetry solution that is compatible with string theory should also come with associated gauge particles. In principle the symmetry breaking process would render them all massive and invisible at low energies. However, we will see that if the symmetry we gauge has mixed anomalies, there is a nice way out of this problem that can lead to a viable axion: the idea is that the symmetry can become an approximate global symmetry of the low energy theory that is only broken by non-perturbative effects. This was first found by Iba\~nez and Quevedo \cite{Ibanez_Quevedo_99} in D-brane models and they argued this way that the proton would be long-lived. This survival of the gauge symmetry as a low energy global symmetry remnant in perturbative effective field theory makes it possible to generate approximate symmetries like those that determine the axion dynamics and hence has observable consequences. This is what we will explore in this paper, in the context of D-brane inspired models. It is important to notice that even though we will be working with string inspired models, the arguments we give can be applied in more general setups and can be argued completely within effective field theory. The string putative embedding restricts the choices of examples we work with giving a simplified set of rules to build models, as well as specific predictions within the models that can be compared to experimental and observational bounds.
Many of the models we explore will be ruled out this way, but not all of them.  More general effective field theory models built with similar rules will have many more free parameters related to mixings that are forbidden in brane models, as well as charges not being quantized by simple rules and are much less predictive.

Let us begin our discussion with having a flavor symmetry that is gauged to address the problem of flavor. This is not a new idea.
The Froggatt-Nielsen mechanism~\cite{Froggatt_Nielsen_79} provides a way to generate the hierarchy of masses while maintaining $\mathcal{O}(1)$ Yukawa couplings, by adding a gauged flavor symmetry to forbid some tree-level fermion masses. This symmetry is spontaneously broken, and effective nonrenormalizable operators of various dimensions provide fermion mass terms. These operators contain different powers of the order parameter for symmetry breaking, which in turns generates hierarchies between fermion masses.

In this paper we make use of this mechanism in the context of simple string-inspired models. This paper is a natural extension of the models introduced in \cite{Berenstein_Perkins_10}. The abelian Froggatt-Nielsen gauge symmetries in these models can have mixed anomalies, which are then cancelled by a generalized Green-Schwarz mechanism \cite{Green_Schwarz_84}. This requires the introduction of a massless pseudoscalar boson that couples to the anomaly and cancels it. Thus, these particles couple to the Standard model like axions and we will call them axions. When these axions cure the anomalies they end up providing St\"uckelberg mass terms for the anomalous gauge bosons. Thus they are eaten up and disappear from the low energy effective theory. However, it turns out that  in D-brane models the global parts of these gauge symmetries survive in the low-energy theory~\cite{Ibanez_Quevedo_99}; we will refer to this phenomenon as the Iba\~nez-Quevedo mechanism. It should be noted that typically, all symmetries in string models should be gauged, and finding systems with approximate global symmetries usually requires unnatural fine tuning; the Iba\~nez-Quevedo mechanism provides a neat way to avoid this. In order to generate fermion mass terms in our models, these residual approximate global symmetries from the Iba\~nez-Quevedo mechanism must be broken. We identify these residual global symmetries as Peccei-Quinn symmetries. 

The Peccei-Quinn symmetry is a classic mechanism which addresses the smallness of the QCD vacuum angle \cite{Peccei_Quinn_77}. One of its low energy consequences is an axion whose properties can be predicted  \cite{Weinberg_78,Wilczek_78}. The simplest types of axion have been ruled out, meaning that the Peccei-Quinn symmetry has to be broken at scales much higher than the electroweak scale, producing an invisible axion. Experimentally, there is a narrow `axion window' which constrains the decay constants of these particles to lie between $10^{9}$ and $10^{12}$ GeV. These lower and upper bounds on the scale of Peccei-Quinn symmetry breaking arise respectively from astrophysical and cosmological considerations; this is reviewed in \cite{Turner_89}. 

Axions are a generic feature of stringy models, usually arising from closed string modes in the context of the Green-Schwarz mechanism. Typically, their decay constants are close to the string scale. Finding realistic axions in stringy models is not always easy, but one can not rule them out either \cite{Svrcek_Witten_06}. Our models include axions of this type, but we include scalars arising from open string modes to break the global Peccei-Quinn symmetry, which induces additional axions. It is these `open string axions' which survive as the physical axion at the lowest energies in our models.
This feature has been found in some top down compactifications  leading to supersymmetric models previously \cite{Kiritsis1}  (these models have been analyzed further in \cite{Kiritsis2}).

The precise string inspiration for this model comes from studying intersecting D-brane setups (see \cite{Blumenhagen_et_al_05} for a review). The matter content of these models is usually presented as a quiver (or moose) diagram. We will follow these conventions. To simplify the analysis we will insist on requiring only the minimal particle content that is able to accommodate the flavor hierarchy generation mechanism at the level of effective field theory. The model we write is
closely related to the `Minimal Quiver Standard Model' introduced in~\cite{Berenstein_Pinansky_07}. The effective field theory will have high dimension operators suppressed by some high scale $M$. For simplicity we will call such scale the string scale, although the model does not need to have a precise string theory origin. We find that if the neutrino masses are included in the analysis, we are able to determine both $M$ and the Froggatt-Nielsen (Peccei-Quinn) symmetry breaking scale, and this gives a prediction of the energy scale of the axion decay constant. Furthermore, we find that in one of the models the axion decay constant falls exactly in the allowed window, while other models can be ruled out because the corresponding axions are not allowed or there are some other basic problems with their phenomenology.

Related work along these lines has been done. In~\cite{ Coriano_Irges_06} a similar scheme of incorporating axions into a Froggatt-Nielsen mechanism is discussed. The authors use a rather different device to break the PQ symmetry in a supersymmetric setting, resulting in Higgs-axion mixing and axion masses and couplings which are not strongly determined by the axion decay constant. A similar result occurs in the models studied in \cite{Kiritsis1, Kiritsis2}. 
These features are not exhibited by the physical axion appearing in our models. Also, our gauged flavor symmetries are abelian, but models which gauge nonabelian flavor symmetries can also yield axionic degrees of freedom, as in~\cite{Berezhiani_83,Berezhiani_Khlopov_91}.

\section{Minimal Models }
\label{S:Quiver}

As we discussed in the introduction, we want a D-brane inspired model that contains the standard model and some gauged flavor symmetries.
Considerations of orientifolded D-brane embeddings of the theory require that all $SU(N)$ gauge symmetries be enhanced to $U(N)$ symmetries at the string scale, but also make the inclusion of $Sp(N)$ gauge symmetries natural. Thus, we require a $U(3)$ color symmetry, but are also allowed to use $Sp(1)\simeq SU(2)$ rather than enhancing isospin symmetry to $U(2)$, avoiding a possible extra $U(1)$ gauge symmetry.
 Having a realistic model with minimal matter content requires an extra brane for a gauged $U(1)$ symmetry, so that the theory has at least a $U(1)\times U(1)$ gauge symmetry. This is true regardless of whether we have $SU(2)$ or $U(2)$ associated to the weak interactions. Although this could in principle be avoided by using models that arise from a grand unified theory, or that contain quarks in a two index antisymmetric representation of $U(3)$, they are extremely problematic as many important couplings of the standard model would need to be generated non-perturbatively ( see \cite{Berenstein:2006aj} for a discussion about this issue).

The minimal matter content would have us select an $SU(2)$ symmetry.
 As is typical of such setups, only hypercharge is a non-anomalous gauged symmetry. The other $U(1)$ symmetry has mixed anomalies that are cancelled by the four dimensional version of the Green-Schwarz mechanism \cite{Green_Schwarz_84}. This produces a heavy $Z'$, and it is possible to generate all couplings in the standard model at tree level, so that the only new physics is the presence of a heavy $Z'$. These considerations produce the minimal quiver model  \cite{Berenstein_Pinansky_07}.

Now, we want to make the model richer by adding a gauged flavor symmetry that distinguishes various quarks from each other. The simplest such extension requires only one more $U(1)$ gauge field. One way to include an extra gauge field is to add another brane to the minimal quiver model, giving a four stack model rather than a three stack model (here stacks also refer to nodes in the quiver diagram). The gauge anomaly cancellation condition $\tr [T^a\{ T^b,T^c \} ] = 0$ can be used to fit the right handed fermions of the standard model with two up-type quarks and one down-type quark at one node, and the opposite arrangement at the other node. The second possibility which we will study further below is to use a $U(2)$ gauge theory for the weak interactions, rather than $SU(2)$. This second choice produces a three stack model instead.

At this stage, we need to fit the Higgs doublet. There are two choices we can make: we can have a model with a Higgs field associated to one $U(1)$ node but not the other; either choice forbids Yukawa couplings for three of the quarks, and possibly all of the leptons.
The three quarks for which Yukawa couplings are allowed are chosen to be the heaviest quarks with the correct quantum numbers. This is, either $t,c,b$ or $t,b,s$. We will call these two choices model A and model B.  This is where we need to add more fields to generate masses for the other generations.
To communicate the electroweak symmetry breaking to the rest of the fermions that connect to the node  where the Higgs field is absent, we introduce a new complex scalar field $\phi$ between the two $U(1)$ nodes that is neutral under the standard model. Another possibility would ask us to have a two Higgs field model instead, one for each $U(1)$ stack: such models besides having more particles and not being minimal would also typically lead to unacceptably large flavor changing neutral currents. Under our assumptions of minimality, we choose to have only a single complex scalar field joining the two $U(1)$ stacks.
We will call the $\phi$ complex scalar the Froggatt-Nielsen (FN) scalar.

\begin{table}[ht]
\centering
\begin{tabular}{|c|c|c|c|c|c|c|c|c|}
\hline
& $\bar u_R^{1,2}$ & $\bar d_R^1$ & $\bar u_R^3$ & $\bar d^{2,3}_R$ & $e_L$ & $\bar e_R$ &$\phi$& $h^{A,B}$ \\
\hline $U(1)_a$ & +1 & -1 & 0 & 0 & 0 & -1 & +1 & +1, 0\\
$U(1)_b$ & 0 & 0 & +1 & -1 & +1 & -1 & -1 & 0, +1\\
\hline
\end{tabular}
\caption{Charges under the Froggatt Nielsen symmetry in the four-stack models. For the Higgs, we have two choices: we call the two models $A,B$}\label{t:one}
\end{table}

In the full model a $U(1)_B\times U(1)_a\times U(1)_b$ is gauged. Hypercharge $U(1)_Y$ is the only anomaly-free linear combination of these. Only the right handed quarks and leptons are charged under $U(1)_{a,b}$, which is the set of gauge groups under which the scalar $\phi$ is charged. Charges under these $U(1)$s are depicted in table \ref{t:one}. Figure~\ref{F:quiver} shows the matter content in quiver form, with both choices for the (single) Higgs doublet present. In figure~\ref{F:quiver} we have left the edges corresponding to right-handed quarks unlabeled since their interpretation as light or heavy depends on which Higgsing scheme is used; generically, up-type right-handed quarks are represented by {\psset{unit=1pt}\begin{pspicture}(20,10)\psline[arrowsize=5]{>->}(0,4)(20,4)\end{pspicture}} edges, while down-type quarks are represented by {\psset{unit=1pt}\begin{pspicture}(20,10)\psline[arrowsize=5]{<->}(0,4)(20,4)\end{pspicture}} edges in the figure. Gauge invariant Yukawa or FN couplings are constucted by tracing closed loops in the diagram. Some of these are high dimension operators. The neutrino masses are also generated in this way, and depending on the choices that are made, they give rise to dimension 5 or dimension 7 operators if they require the introduction of FN fields to render the couplings that generate masses gauge invariant.

\begin{figure}
\centering
\psset{unit=5mm}
\begin{pspicture}(10,10)
\cnodeput(0,9){A}{$U(1)_{a}$}
\cnodeput(9,9){B}{$U(1)_{b}$}
\cnodeput(5,6){C}{$U(3)_{c}$}
\cnodeput(5,0){W}{$Sp(1)_{w}$}
\psset{nodesep=0mm}
\ncline[arrowsize=2pt 5]{>-}{C}{W}
\psset{nodesep=2mm}
\ncline[arrowsize=2pt 5]{>-}{C}{W}
\psset{nodesep=4mm}
\ncline[arrowsize=2pt 5]{>-}{C}{W}
\Bput*{$q_{i}$}
\psset{nodesep=0mm}
\ncarc[linestyle=dashed,arrowsize=2pt 5,arcangle=35]{>->}{A}{B}
\Aput*{$\phi$}
\psset{nodesep=0mm}
\ncarc[arrowsize=2pt 5,linestyle=dotted,arcangle=30]{>-}{B}{W}
\Aput*{$h^{B}$}
\psset{nodesep=0mm}
\ncarc[arrowsize=2pt 5,linestyle=dotted,arcangle=30]{-<}{W}{A}
\Aput*{$h^{A}$}
\psset{nodesep=0mm}
\ncarc[arrowsize=2pt 5,arcangle=-5]{<->}{B}{A}
\psset{nodesep=2mm}
\ncarc[arrowsize=2pt 5,arcangle=-5]{<->}{B}{A}
\psset{nodesep=4mm}
\ncarc[arrowsize=2pt 5,arcangle=-5]{<->}{B}{A}
\Bput*{$\bar{e}_{i}$}
\psset{nodesep=0mm}
\ncarc[arrowsize=2pt 5,arcangle=-10]{-<}{W}{B}
\psset{nodesep=2mm}
\ncarc[arrowsize=2pt 5,arcangle=-10]{-<}{W}{B}
\psset{nodesep=4mm}
\ncarc[arrowsize=2pt 5,arcangle=-10]{-<}{W}{B}
\Aput*{$l_{i}$}
\psset{nodesep=0mm}
\ncarc[arrowsize=2pt 5]{<->}{C}{A}
\Aput*{$\bar{D}$}
\psset{nodesep=0mm}
\ncarc[arrowsize=2pt 5]{>->}{A}{C}
\psset{nodesep=2mm}
\ncarc[arrowsize=2pt 5]{>->}{A}{C}
\Aput*{$\bar{u}_{I}$}
\psset{nodesep=0mm}
\ncarc[arrowsize=2pt 5]{>->}{B}{C}
\Aput*{$\bar{U}$}
\psset{nodesep=0mm}
\ncarc[arrowsize=2pt 5]{<->}{C}{B}
\psset{nodesep=2mm}
\ncarc[arrowsize=2pt 5]{<->}{C}{B}
\Aput*{$\bar{d}_{I}$}
\end{pspicture}
\vspace{5mm}
\caption{Quiver with extra scalar $\phi$. The two choices for the Higgs doublet are labeled by the dotted lines and the model they represent $h^{A,B}$ (for the models A,B, respectively). We have added arrows to indicate the $U(1)_{ab}$ charges of $\phi$.}\label{F:quiver}
\end{figure}
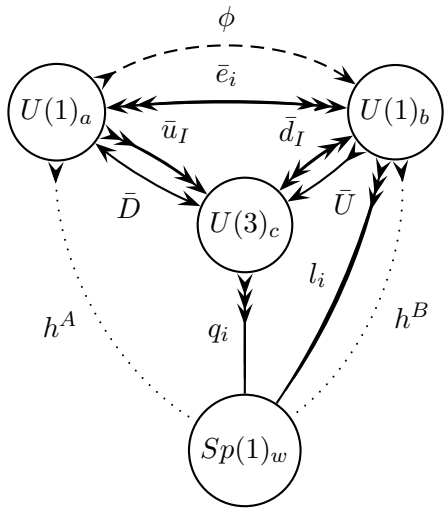

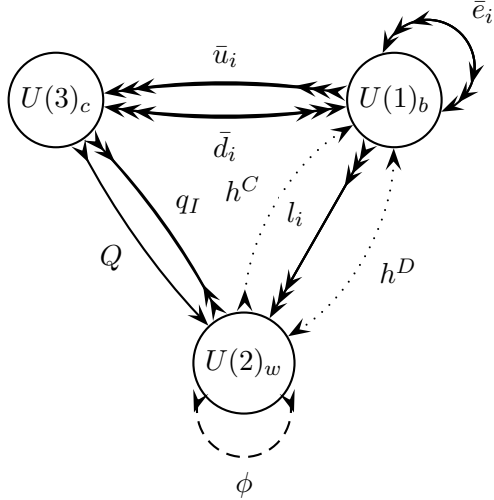
\begin{figure}[ht]
\vskip 1 cm
\centering
\psset{unit=5mm}
\begin{pspicture}(10,10)
\cnodeput(9,9){B}{$U(1)_{b}$}
\cnodeput(0,9){C}{$U(3)_{c}$}
\cnodeput(5,2){W}{$U(2)_{w}$}
\psset{nodesep=0mm}
\ncarc[arrowsize=2pt 5]{<-<}{W}{C}
\Aput*{$Q$}
\psset{nodesep=0mm}
\ncarc[arrowsize=2pt 5]{>-<}{C}{W}
\psset{nodesep=2mm}
\ncarc[arrowsize=2pt 5]{>-<}{C}{W}
\Aput*{$q_{I}$}
\psset{nodesep=0mm}
\ncarc[linestyle=dotted,arcangle=-29,arrowsize=2pt 5]{<-<}{B}{W}
\Bput*{$h^{C}$}
\psset{nodesep=0mm}
\ncarc[linestyle=dotted,arcangle=29,arrowsize=2pt 5]{<->}{B}{W}
\Aput*{$h^{D}$}
\psset{nodesep=0mm}
\ncarc[arrowsize=2pt 5,arcangle=0]{<-<}{W}{B}
\psset{nodesep=2mm}
\ncarc[arrowsize=2pt 5,arcangle=0]{<-<}{W}{B}
\psset{nodesep=4mm}
\ncarc[arrowsize=2pt 5,arcangle=0]{<-<}{W}{B}
\Aput*{$l_{i}$}
\psset{nodesep=0mm}
\ncarc[arrowsize=2pt 5]{<-<}{C}{B}
\psset{nodesep=2mm}
\ncarc[arrowsize=2pt 5]{<-<}{C}{B}
\psset{nodesep=4mm}
\ncarc[arrowsize=2pt 5]{<-<}{C}{B}
\Aput*{$\bar{u}_{i}$}
\psset{nodesep=0mm}
\ncarc[arrowsize=2pt 5]{<->}{B}{C}
\psset{nodesep=2mm}
\ncarc[arrowsize=2pt 5]{<->}{B}{C}
\psset{nodesep=4mm}
\ncarc[arrowsize=2pt 5]{<->}{B}{C}
\Aput*{$\bar{d}_{i}$}
\psset{nodesep=0mm}
\nccircle[angleA=180,linestyle=dashed,arrowsize=2pt 5]{>-<}{W}{1.25}
\Bput*{$\phi$}
\psset{nodesep=0mm}
\nccircle[angleA=315,arrowsize=2pt 5]{<->}{B}{1.25}
\psset{nodesep=2mm}
\nccircle[angleA=315,arrowsize=2pt 5]{<->}{B}{1.25}
\psset{nodesep=4mm}
\nccircle[angleA=315,arrowsize=2pt 5]{<->}{B}{1.25}
\Bput*{$\bar{e}_{i}$}
\end{pspicture}
\vspace{5mm}
\caption{A three-stack model yielding FN terms. Both choices for Higgs charges, models C and D, are depicted in the diagram.}\label{F:threestack}
\end{figure}

A mass hierarchy from a FN symmetry can also be introduced in a three-stack model, once again by a slight modification of the minimal quiver model. This is done by replacing $Sp(1)$ by $U(2)_w$, which introduces another gauged $U(1)$ boson.  Tadpole cancellation requires that the three left-handed quark doublets be split into two sets, with two generations in the fundamental and one in the antifundamental representation of $U(2)_w$. This also ensures that the FN symmetry acts differently on different families as we want for our models. Incidentally, this prevents the $U(1)\subset U(2)$ from giving contributions to the hypercharge $Y$. Fixing the hypercharge of the Higgs doublets requires fixing the arrows of the field at node $U(1)_b$. The choice now becomes on how the Higgs transforms under $U(2)_w$: in the fundamental or the antifundamental representation of $U(2)_w$. As far as the $SU(2)_w$ is concerned, both of these choices are equivalent, but they differ in how the Higgs is charged under the $U(1)_w$.
 This is pictured in figure~\ref{F:threestack}. As in the previous models, there are two ways to charge the Higgs doublet, resulting in two models which we refer to as model C and model D.

Once again, some Yukawa couplings are forbidden, and we include an additional scalar $\phi$ in order to generate FN terms and fully communicate electroweak symmetry breaking. This scalar is chosen to be in the antisymmetric representation of $U(2)$ so that $\phi$ is a singlet under standard model isospin, but remains charged under the extra $U(1)_w\subset U(2)_w$. Left-handed neutrino masses are generated in identical fashions to models A and B: by closing loops in the diagram and paying attention that
the arrows at the nodes $U(1)_b$ and $U(2)_w$ are oriented so that the gauge charges cancel.

\begin{table}[ht]
\centering
\begin{tabular}{|c|c|c|c|c|c|}
\hline
& $ q_{I} $ & $ Q $ & $ l_{i} $ & $ \phi $ & $ h^{C,D} $  \\
\hline $U(1)_w$ & +1 & -1 & -1 & +2 & 1,-1 \\
\hline
\end{tabular}
\caption{Charges under the Froggatt Nielsen symmetry in the three-stack models. For the Higgs, we list the two choices $C,D$\label{t:two}}
\end{table}

Given the quantum numbers we have described, the minimal set of  couplings that generate all Yukawa couplings, including masses for left-handed neutrinos, is shown in tables \ref{T:modelA}-\ref{T:modelD}. The high dimension operators (we will call these the Froggatt-Nielsen terms of the action) are suppressed by some high scale, $M$, which we take to be the string scale; this is a minimality assumption in order to be predictive. The suppression splits the fermions into light and heavy tiers. The heaviest field in each of the light and heavy tiers is taken to have an order one dimensionless coupling to the scalar fields; the ratio between tiers is set by the mass ratio of the heaviest quarks in each tier.  This can be used to get an order of magnitude estimate on both $\vev{\phi}$ and the string scale $M$ by examining mass ratios. We assume that the heaviest neutrino mass saturates the upper bound of $m_{\nu_{\tau}}\simeq 2$ eV, and using the usual Higgs VEV $\vev{h}=246$ GeV we get the estimates given in the tables in section \ref{S:Features}.

Notice that with this procedure we are able to predict all relevant scales for the model from known data. The reason for this is that we have used the neutrino masses to find the scale of new physics, since these are the parameters in the standard model that have been observed that are sensitive to the high scale physics (it arises from non-renormalizable couplings at the electroweak scale in the absence of right handed neutrinos, or when these have been integrated out). If in a given model the neutrino masses have a different origin than the string scale or the Froggatt-Nielsen scale, the above constraints change and tend to push the string scale higher. At this stage we must ask what is the physical implication of these deductions for low energy physics. In particular, we must determine how $\vev{\phi}$ is related to the decay constant of the axion appearing in the low energy theory, since the stringent bounds on this decay constant can be used to rule out models of axions. In order to proceed, we need to examine the effective field theory more carefully. We will do this in two steps. In the next section, section \ref{S:Anomalies}, we describe how to extract the axion in the general class of models we have constructed. In section \ref{S:axionc} we determine the axion  model parameters that are needed to understand the axion couplings to matter.

Notice that our models generate various textures for quark masses. These have been analyzed in a fairly general way in \cite{Kiritsis3} for other models (see also \cite{Cvetic2} for specific examples), but the general form of the textures is very similar.

\section{The physical axion}
\label{S:Anomalies}

Let us begin our discussion with the $U(1)$ gauge fields. We have a kinetic term of the form
\begin{equation}
{\cal L}_{U(1)^3, kin} = \frac{1}{4g_a^2} (F_{\mu\nu}^a)^2 + \frac {1}{4g_b^2}(F^b_{\mu\nu})^2+ \frac {1}{4g_c^2}  (F^c_{\mu\nu})^2\label{eq:kin}
\end{equation}
and there are no mixed terms between the kinetic terms in D-brane models at tree level. These can be generated by radiative corrections, but we will ignore this at this
stage.

The $U(1)_{i}$ charges $Q_{i}$ have mixed $U(1)^{3}$, $U(1)Sp(1)^{2}$, and $U(1)SU(3)^{2}$ anomalies in each of the models. The full anomaly needs to be cancelled in order to have sensible physics.  In each of the four models, we find that one linear combination $Q_{Y}$ of these charges is anomaly-free, and two are anomalous. In all four models, $Q_{Y}$ couples to the anomaly-free gauge field $Y_{\mu}$, which is some linear combination of the original three $U(1)$ fields. We take the two remaining, anomalous charges $\tilde{Q}_{i}$ to be orthogonal to the anomaly-free charge, and to each other.  These linear combinations can be determined by the vanishing of the three-way $U(1)$ anomaly. Knowing all three anomalous and anomaly-free linear combinations allows one to determine the mixing parameters of these gauge fields (analogous to the weak mixing angle $\sin^{2}\theta_{W}$) in terms of the three original couplings $g_{i}$.

Many mixed anomalies remain uncancelled. We have taken care of the $Q_{Y}\tilde{Q}_{1}\tilde{Q}_{2}$ anomaly already, by the precise choice of $Q_1 Q_2$; the rest of the triangle diagrams involving the gauge bosons $\tilde{Z}^{i}$ associated with $\tilde{Q}_{i}$ are cancelled via the Green-Schwarz mechanism. The remaining mixed $U(1)$ anomalies can be cancelled by including Chern-Simons terms in the effective action (see~\cite{Antoniadis_et_al_00} for a discussion). When describing the Green-Schwarz mechanism, the mixed abelian-nonabelian anomalies are cancelled by the introduction of two additional pseudoscalar fields $s^{i}$ with axion-like couplings to the standard model gauge fields. These pseudoscalar fields transform non-linearly under the anomalous $U(1)$ gauge transformations. Their gauge invariant kinetic term is of the form
\begin{equation}
\frac 12 K_{ij} (\partial_\mu s^i - \tilde Z_\mu^i)(\partial_\mu s^j - \tilde Z_\mu^j)
\end{equation}
and the scale of $K$ is close to $M^2$.

Since the scalars transform non-linearly, we can choose the unitary gauge where the $s^{i}$ are constant. In this gauge the $\tilde Z^{i}$ are clearly seen to acquire a  St\"uckelberg mass of order $g M$ , where $g$ is of order $g_a,g_b,g_c$ which are small at high energies (the factor of $g$ is the usual open-string-closed string mixing), so they can be integrated out of the low energy effective theory.  When integrating the $Z'$ fields, this process produces only current-current interactions for the $U(1)$ gauge fields that have been integrated out, but these currents are invariant under the anomalous global $U(1)$s. Starting from this simple action, the low energy effective theory seems to have a $U(1)^2$ global symmetry in the lagrangian, but in this case this symmetry is anomalous. This is a particular realization of the Iba\~nez-Quevedo mechanism \cite{Ibanez_Quevedo_99}. This mechanism is usually used to guarantee a long lifetime for the proton, but it serves generically to produce approximate global symmetries in the infrared effective theory.

One of these symmetries is baryon number (this is used to prevent proton decay), while the other symmetry is the Froggatt-Nielsen symmetry of our lagrangian. The field $\phi$ is then a scalar field that
breaks the Froggatt-Nielsen symmetry, and we get an additional Goldstone boson for this symmetry breaking.
Notice that if the $U(1)_{FN}$ gauge symmetry were non anomalous, this Goldstone boson would have been eaten by the Froggatt-Nielsen gauge field and there would be no low energy remnant. This is because having a St\"uckelberg mass for the gauge field would not be required by anomaly cancellation. Also, since the symmetry that is broken is anomalous in the low energy effective theory, the symmetry is not a true symmetry of the lagrangian anyhow and non-perturbative effects will lift the symmetry producing a potential for the would-be Goldstone boson. Such terms can be traced to the original lagrangian and they require
corrections to the potential of the form
$
V( \phi \exp( i s),\phi^* \exp(-i s) )
$, as described in \cite{Coriano_Irges_06}. These terms are not allowed by perturbative computations in string theory
so they can only be generated by D-brane instantons (see \cite{Blumenhagen_et_al_09} for a review of instanton computations). This means that the would be Goldstone boson is very light and gives rise to a light invisible axion \cite{Kim_79}. If we take the parametrization $\phi=
r \exp(i\theta)$, then $\theta$ is the phase of $\phi$ that has been identified with an axion field at intermediate energy.

A more careful analysis of the lagrangian shows that the field $\theta$ will mix with the $s^i$ fields and that the vev of $\phi$ will also contribute to the mass term of the $\tilde Z$ fields. If the $\tilde Z$ fields have a bare mass generated at the string scale, the new contribution would be a small correction at the Froggatt-Nielsen scale, so the
correction from mixing of scalars is small and $\theta$ plays the role of the intermediate energy axion.
This is what we expect in the limit of a high string scale - to a good approximation the physical axion is just the phase of $\phi$, and the axion decay constant is simply $f_a= \vev{\phi}$.

This can be seen explicitly by considering the St\"uckelberg terms for both the Green-Schwarz axions and for $\theta$, which  cause these three scalars to mix. Rewriting the St\"uckelberg terms for all the scalars in terms of $\tilde{Z}^{1,2}_{\mu},Y_{\mu}$, we find that $Y_{\mu}$ disappears, since $\phi$ is not charged under $Q_{Y}$ in any of the models. Altogether we have

\[
\frac{1}{2}(\partial_{\mu}s^{1} + f_{s^1}g_{z^1}\tilde{Z}^{1}_{\mu})^{2} + \frac{1}{2}(\partial_{\mu}s^{2} + f_{s^2}g_{z^2}\tilde{Z}^{2}_{\mu})^{2} + \frac{1}{2}(\partial_{\mu}\theta + P_{z^1}g_{z^1}\tilde{Z}^{1}_{\mu}\vev{\phi} + P_{z^2}g_{z^2}\tilde{Z}^{2}_{\mu}\vev{\phi})^{2}
\]

We see that the Goldstone boson mixes with the Green-Schwarz axions to form new scalars
\[
\alpha_{i}
\equiv
\frac{f_{s^i}g_{z^i}}{\mu_{z^i}}s^i  + \frac{\vev{\phi} P_{z^i}}{\mu_{z^i}}\theta
\sim
\frac{s^i + \varepsilon\theta}{\sqrt{1 + \varepsilon^2}}
\]
for mixing parameters and couplings $P_{z^i}$ and $\mu_{z^i}\equiv \sqrt{(f_{s^i}g_{z^i})^{2}+(\vev{\phi} P_{z^i})^2} \sim M\sqrt{1 + \varepsilon^2}$. In the approximations we take $P_{z^i} \sim g_{z^i} \sim \mathcal{O}(1)$, $f_{s^i} \sim M$, and $\varepsilon \equiv \vev{\phi}/M \ll 1$. These axions $\alpha_{i}$ are eaten by the heavy gauge bosons $\tilde{Z}^{i}_{\mu}$. The heavy gauge bosons acquire the masses $M_{i} \simeq \mu_{z^{i}}$ from eating these scalars. The combination of $s^{i}, \theta$ orthogonal to $\alpha_{i}$ is the physical axion $a$, surviving in the low-energy theory:

\begin{align*}
a
&\equiv
\frac{
(-f_{s^{2}}g_{z^{2}}\vev{\phi} P_{z^{1}})s^{1} +
(-f_{s^{1}}g_{z^{1}}\vev{\phi} P_{z^{2}})s^{2} + (f_{s^{1}}g_{z^{1}}f_{s^{2}}g_{z^{2}})\theta} {\sqrt{(\mu_{z^{1}}\mu_{z^{2}})^{2}-\vev{\phi}^{4}(P_{z^{1}}P_{z^{2}})^{2}}} \\
& \sim
\frac{\theta - \varepsilon(s^1 + s^2)}{ \sqrt{ (1+\varepsilon^2)^2 - \varepsilon^4 } }
\sim
\theta - \varepsilon(s^1+s^2)
\end{align*}

So indeed, up to corrections of order $\vev{\phi}/M$, the physical axion is just the phase of $\phi$. Since the Higgs is also charged under the Froggatt-Nielsen symmetry, when the electroweak theory is broken, there is in principle a further mixing between $\phi$ and the neutral would-be Goldstone boson from the Higgs doublet. However, at the level of the effective action at the weak scale no such terms seem to appear in the effective potential in our models. Any such mixing would be produced from mixing with the high scale $\tilde Z'$ and is very suppressed, by $v^2/M^2$, so it can be safely ignored.

The $W^{\pm}$ fields are as in the standard model, but $Z$ differs somewhat. In models A and B, the situation is as follows; that for models C and D is analogous. Representing the Higgs doublet as $h=(\langle h\rangle+H+iz,\,w^{+})$, we find the St\"uckelberg term
\[
\left(\partial_{\mu}z + g_{w}W^{3}_{\mu}\vev{h} + 2g_{b}\vev{h}(P_{bz^1}\tilde{Z^{1}}_{\mu} + P_{bz^2}\tilde{Z^{2}}_{\mu} + P_{by}Y_{\mu})\right)^{2}
\]
upon writing $A^b_{\mu}$ in the $Y\tilde{Z}^{1,2}$ basis. We can determine the mixing parameter $P_{by}$ by comparing the abelian parts of the gauge covariant derivatives in the $A^{a,b,c}$ and $Y\tilde{Z}^{i}$ bases, using the known relation between $U(1)$ charges $Q_{Y}=\frac{1}{6}(Q_{c}-3Q_{a}-3Q_{b})$, and the fact that the mixing matrix $P_{ij}$ is orthogonal. We find that $g_{b}A^b_{\mu} = -\frac{1}{2}g_{y}Y_{\mu} + \ldots$, so that the relevant part of the covariant derivative is

\begin{align*}
g_{w}W^{3}_{\mu}T^{3} + g_{b}A^b_{\mu}Q_{b}
& =
(g_{w}W^{3}_{\mu}-g_{y}Y_{\mu}) + 2g_{b}(P_{bz^1}\tilde{Z}^{1}_{\mu} + P_{bz^2}\tilde{Z}^{2}_{\mu}) \\
& \equiv \tilde{M}_{Z}Z_{\mu} + \tilde{M}_{z^1}\tilde{Z}^{1}_{\mu} + \tilde{M}_{z^2}\tilde{Z}^{2}_{\mu}
\end{align*}

The first term in parentheses in the center is the usual $Z$ boson, and the other two terms form some combination of $\tilde{Z}^{i}$ bosons. None of the St\"uckelberg terms left after EWSB involve any mixing between $a$ and any of the Goldstone modes from the Higgs (but there is mixing of these with the other `axions' $\alpha_{i}$). Thus, this model does not provide an `axi-Higgs,' as in \emph{e.g.}~\cite{Coriano_Irges_06}.

Since axions are so light, and can in principle be produced in profusion in low-energy nuclear processes, they must not couple strongly enough to nuclear matter to quench stellar fusion. This is the source of the lower constraint on the axion decay constant. This bound is model-dependent, so it is useful to check that the nominal axion window holds for the models under consideration here. In particular we wish to determine whether there are low-energy couplings which accidentally lead to fine-tuning and radically move the lower sill of the axion window in these models.

\section{Axion Couplings}
\label{S:axionc}

In order to determine the couplings of the axion to matter at low energies, we follow the procedure of~\cite{Georgi_et_al_86}, which involves a series of matter field redefinitions. We perform two phase rotations of the matter fields, one to realize the PQ symmetry nonlinearly for physics below the PQ scale, and the other to eliminate axion-gluon couplings for physics below the QCD scale. These field redefinitions result in derivative couplings of the axion to currents, as is characteristic of Goldstone bosons. These phase rotations are also anomalous, resulting in axion-field strength couplings in the effective lagrangian. At the very lowest energies, we use these couplings to place bounds on the axion decay constant from nuclear processes.

The first field redefinition is performed below the $PQ$ scale, where in our models, the extra $U(1)$ gauge symmetries are broken, and only the standard model gauge group remains. In this field redefinition, we rotate the phase of $\phi$ onto the fermion fields in the FN terms, making use of the unbroken PQ symmetry:

\[
\phi\psi\overline{\psi}h\rightarrow \left(e^{ia X_{\phi}/f}\phi\right) \left(e^{ia X_{\psi}/f}\psi\right) \left(e^{ia X_{\overline{\psi}}/f}\overline{\psi}\right)h
\]
where $\psi,\overline{\psi}$ are generic left-handed Weyl fields; the bar indicates an antiparticle. We require the PQ charges $X_{i}$ to sum to $-1$ in each FN term in order to cancel the phase of $\phi$. For now, we leave these charges unspecified; we will discuss the choice of these charges below.

This results in the coupling of the axion to the PQ current

\[
\frac{\partial_{\mu}a}{f_{a}}\; \overline{\Psi}\gamma^{\mu}\left(\frac{X_{\psi}-X_{\overline{\psi}}}{2} - \frac{X_{\psi}+X_{\overline{\psi}}}{2}\;\gamma_{5}\right)\Psi
\]

The vector part of this coupling violates CP - we will discuss a resolution below. Since the PQ transformation is anomalous, it induces the axion-gauge interaction terms

\begin{align*}
\mathcal{L}_{a}
& =
- \frac{1}{2} \displaystyle\sum_{i}\frac{g_{i}^{2}}{8\pi^{2}f_{a}} a\;\tr[Q_{PQ}T_{i}^{a}T_{i}^{b}] F_{i}^{a}\tilde{F}_{i}^{b} \\
& \equiv
- \frac{a}{16\pi^{2}f_{a}}
\left( g_{c}^{2}N G\tilde{G} + g_{w}^{2} C_{W} W\tilde{W} + g_{Y}^{2} C_{Y} Y\tilde{Y} \right)
\end{align*}
where the sum runs over the standard model gauge groups, and the trace runs over the fermion fields coupling to each. The extra overall factor of $1/2$ arises because the field rotations are chiral. We have defined the PQ color anomaly $N = \frac{1}{2}\tr[Q_{PQ}]$. Below the weak scale the hypercharge and weak anomalies merge to give the electromagnetic anomaly coefficient $E \equiv C_{W} + C_{Y} =\tr[Q_{PQ}Q_{E}Q_{E}] $, so that only axion couplings to gluons and photons remain.

The second field redefinition is performed below the weak scale, in order to eliminate the axion-gluon coupling in favor of couplings of axions to light quarks. In a sense this is the reverse of the first field redefinition, except that here we only rotate the light quark fields

\[
\Psi_{q} \rightarrow e^{iaQ_{R}(2N)/f_{a}}\Psi_{q}
\]

The charge $Q_{R} = Q_{V}+Q_{A}\gamma_{5}$ is anomalous, so its axial part modifies the effective axion-gauge interactions to

\[
-\frac{g_{c}^{2}}{8\pi^{2}f_{a}}\,
\left(\frac{1}{2}N - 2N \frac{1}{2} \tr[Q_{A}]\right)
\,a G^{a}\tilde{G}^{b}
- \frac{e^{2}}{8\pi^{2}f_{a}}\,
(\frac{1}{2}E - 2N\tr\left[Q_{A}Q_{E}Q_{E}\right])
\,a F\tilde{F} \\
\]
with $Q_{E}$ the electric charge matrix of the light quarks. We need $\tr\, Q_{A}=1/2$ in order to cancel away the axion-gluon coupling, but the choice of $Q_{R}$ is otherwise unrestricted. Convenient choices can be made to simplify various interactions. The light quark interactions are also modified to

\[
\frac{\partial_{\mu}a}{f_{a}}\; \overline{\Psi}\gamma^{\mu}\left\{ \left(\frac{X_{\psi}-X_{\overline{\psi}}}{2} + Q_{V}\right) - \left(\frac{X_{\psi}+X_{\overline{\psi}}}{2}- Q_{A}\right)\;\gamma_{5}\right\}\Psi \\
\]

The axial part of this current contributes to the axion-hadron couplings. The axion-baryon couplings are determined via current algebra in~\cite{Kaplan_85} to be

\[
\frac{\partial_{\mu}a}{f_{a}}\left\{ 2\,(\tr \mathcal{Q}_{A}T^{a}) \left(F\,\tr\overline{B}\gamma^{\mu}\gamma_{5}[T^{a},B] + D\,\tr\overline{B}\gamma^{\mu}\gamma_{5}\{T^{a},B\}\right) + \frac{S}{3}(\tr \mathcal{Q}_{A})\tr\overline{B}\gamma^{\mu}\gamma_{5}B \right\}
\]
where $B$ is the baryon octet matrix, $\mathcal{Q}_{A} = \frac{X_{\psi}+X_{\overline{\psi}}}{2}-Q_{A}$, and $F=0.44$, $D=0.81$, $S=0.13$ are baryon decay parameters~\cite{PDG,Kim_C_08}. This gives a derivative pseudovector coupling $g_{aNN}(\partial_{\mu}a/f_{a}) \bar{N}\gamma^{\mu}\gamma_{5}N$ to nucleons.

It remains to specify the charges $Q_{PQ}, Q_{R}$ for each model. One might hope that this could be done in such a way as to cancel off the vector part of the axion current, since it seems to be a source of additional CP violation in the strong sector, and complicates the calculation of bounds on the axion decay constant. This is not possible in general, because of the restrictions on the choice of the PQ charges. In these models the PQ symmetry is the global remnant of a gauge symmetry, so it must be some linear combination of the three $U(1)$ generators in the quiver; that is, $X_{\psi} = \sum a_{i}Q_{i}(\psi)$. Given the $U(1)$ charges of the light quarks, and the fact that $Q_{A}$ must trace to $1/2$, we find that it is not possible in any of the models to find a choice of $a_{i}$ and $Q_{V}$ leading to $(X_{\psi}-X_{\overline{\psi}})/2 + Q_{V}=0$ for each light quark.

The vector part of the axion current can, however, be redefined away by yet another chiral rotation of the quark fields, done independently on each light quark. This has the effect of absorbing the offending CP violation into CKM matrix elements, so the seeming strong CP violation is really a weak CP violation. This might seem to violate electroweak symmetry, since up and down type quarks in the same isospin doublet are transformed differently, but this transformation is done in an effective theory at a scale where the electroweak symmetry is already broken.

Given this, we choose $Q_{PQ}$ and $Q_{R}$ to simplify calculations, and determine the couplings of the axion to matter in each model. Some of the Yukawa couplings must be somewhat finely tuned in order to reproduce the observed mass hierarchy. $Q_{A}$ is chosen to simplify hadronic axion interactions as much as possible. In general, axions suffer mixing with $\pi^{0},\eta$. Mass mixing can be eliminated by choosing

\[
Q_{A} = \frac{M_{q}^{-1}}{2\tr M_{q}^{-1}} = \frac{1}{2(1+z+w)}
\left(\begin{array}{ccc}
1 & & \\
& z & \\
& & w \\
\end{array}\right)
\]
with $M_{q}$ the light quark mass matrix and $z=m_{u}/m_{d}$, $w=m_{u}/m_{s}$. This choice of $Q_{A}$ can be seen as either eliminating axion-meson bilinears in an effective meson lagrangian including axions~\cite{Georgi_et_al_86}, or directly via current algebra as eliminating off-diagonal terms in the Dashen formula for meson masses~\cite{Srednicki_85,Dashen_69}. The resulting axion mass is

\[
m_{a}^{2} = \frac{4m_{\pi}^{2}f_{\pi}^{2}z}{(1+z)(1+z+w)f_{a}^{2}} \\
\]

If we also make the choice $Q_{A} = Q_{V}$ then $Q_{R}$ is purely right chiral, which simplifies the axion weak interactions. The Higgs is uncharged under the PQ symmetry, thus preventing axion-Higgs mixing. There is still kinetic axion-meson mixing, but these cross terms are suppressed by $f_{\pi}/f_{a}$. For each model, $Q_{PQ}$ is chosen so that $\tr[Q_{PQ}] = 1$. In each case, the quiver structure allow us to choose $Q_{PQ}$ to be a very simple linear combination of the anomalous $U(1)$ charges; the specific choices for each model are described in the next section.

We can now determine the axion-photon couplings. With the above choice of $Q_{A}$, the axion-photon interaction term becomes

\[
- \frac{e^{2} N }{16\pi^{2}f_{a}}\,
\left(\frac{E}{N} - \frac{2}{3}\frac{4+z+w}{1+z+w}\right)
\,a F\tilde{F} \equiv g_{a\gamma}\, a \mathbf{E\cdot B} \\
\]

In all our models $N=1/2$, and $E$ will be listed in the next section.

\section{Model features}
\label{S:Features}

\subsection{Four-Stack Models}

\subsubsection{Model A}

Model A has the most natural set of FN couplings for quarks; since the $u,d,s$ quarks have suppressed masses in this model, the FN couplings must be down by only $10^{-2}$ at worst for the up and charm quarks. This does not work for the leptons, which naturally acquire masses near the top quark mass in this model. We consider the possibility of fine tuning at this stage (assuming that there is some other effect making these couplings small that has not been included in the model because of our minimality assumptions) and will check the axion phenomenology to see the features that show up. For this model, we choose $Q_{PQ}=-Q_{b}$.

The linear combination $Q_{Y}=\frac{1}{6}\left(Q_{c} - 3Q_{a} - 3Q_{b}\right)$ is anomaly-free, and corresponds to the electroweak hypercharge. This anomaly-free charge must be orthogonal to the two anomalous charges, which constrains them to have the form $\tilde{Q}_{i} = (\gamma_{i}/3-\beta_{i})Q_{a}+\beta_{i} Q_{b} + \gamma_{i} Q_{c}$. We wish to determine $\beta_{i},\gamma_{i}$. First, we require the three-way $U(1)$ anomaly coefficient to vanish; for model A, this is $\tr[Q_{Y}\tilde{Q}_{1}\tilde{Q}_{2}] = \frac{10}{3}\gamma_{1}\gamma_{2} - 6\beta_{1}\beta_{2}$. We also demand that the anomalous charges be orthogonal to each other. The overall normalization of the anomalous charges is arbitrary, so finally we may choose $\gamma_{i}=1$. This gives us two equations for the two unknowns $\beta_{i}$

\begin{align*}
\tr[Q_{Y}\tilde{Q}_{1}\tilde{Q}_{2}]
& =
\frac{10}{3} - 6\beta_{1}\beta_{2} = 0\\
\tilde{Q}_{1}\cdot\tilde{Q}_{2}
& = \left(\frac{1}{3}-\beta_{1}\right)\left(\frac{1}{3}-\beta_{2}\right) + \beta_{1}\beta_{2} + 1 = 0
\end{align*}

which gives us the anomalous charges

\begin{align*}
\tilde{Q}_{1} = \frac{-9+\sqrt{95}}{3}Q_{a} - \frac{10-\sqrt{95}}{3}Q_{b} + Q_{c},
\qquad
\tilde{Q}_{2} = \frac{5-\sqrt{95}}{5}Q_{a} - Q_{b} + \frac{3(10-\sqrt{95})}{5}Q_{c} \\
\end{align*}

The following fermions receive their masses from dimension $4$ operators: $t, b, c, \ell$. To accommodate the neutrino masses we find that we need dimension seven operators in the effective field theory. This is due to the combinations of the FN charge of the lepton doublets and the Higgs.  The results are summarized in the table
\ref{T:modelA}.

\begin{table}
\centering
\begin{tabular}{|c|c|}
\hline
Dimension 4 couplings & $h^{\dagger} q \bar{u}_{I} + h q \bar{D} + h l\bar{e}$\\
\hline
High dimension couplings & $\phi h^{\dagger} q\bar{U} + \phi^{\dagger} h q\bar{d}_{I} + (\phi h^{\dagger}l)^{2}$\\
\hline
$\vev{\phi}\simeq\frac{m_{s}}{m_{t}}\frac{m_{s}^{2}}{m_{\nu_{\tau}}}$ & $10^{3}$ GeV \\
\hline
$M\simeq\frac{m_{s}^{2}}{m_{\nu_{\tau}}}$ & $10^{6}$ GeV\\
\hline
$g_{app}$ & 0.92\\
\hline
$g_{ann}$ & -0.58\\
\hline
$E$ & $-13/12$\\
\hline
$g_{a\gamma}$ & $-5\times 10^{-6}$ GeV$^{-1}$\\
\hline
\end{tabular}
\caption{Model A}\label{T:modelA}
\end{table}

The most important thing to notice is that the since the neutrino masses come from dimension seven operators means that the neutrino masses require the high energy scale to be low
From table~\ref{T:modelA} we see that none of the axion couplings are unusually small, so the axion window is not shifted from its nominal range. The axion in model A is ruled out, since its decay constant falls many orders of magnitude below the axion window.

\subsubsection{Model B}

Model B fares much better than model A. It predicts light $u,d,c$ quarks and light leptons; this puts the smallest couplings in each tier down by $10^{-3}$. Only the Yukawa coupling of the strange quark is unnaturally small. Since this model differs only in its Higgs sector from model A, the anomaly-free hypercharge and anomalous $U(1)$ charges are identical to those in model A. For this model, we choose $Q_{PQ}=Q_{a}$.

Again, none of the couplings to matter turn out to be unusually small. Model B produces a harmless axion, since its decay constant falls squarely in the center of the allowed window, as indicated in table~\ref{T:modelB}. Model B would have been preferred anyhow because in that model the leptons have masses comparable to light quarks without fine tuning.

\begin{table}
\centering
\begin{tabular}{|c|c|}
\hline
Dimension 4 couplings & $h^{\dagger} q \bar{U} + h q \bar{d}_{I}$\\
\hline
High dimension couplings & $\phi^{\dagger} h^{\dagger} q\bar{u}_{I}+ \phi h q\bar{D} + \phi hl\bar{e} + (h^{\dagger}l)^{2}$\\
\hline
$\vev{\phi}\simeq\frac{m_{c}}{m_{t}}\frac{m_{t}^{2}}{m_{\nu_{\tau}}}$ & $10^{11}$ GeV \\
\hline
$M\simeq\frac{m_{t}^{2}}{m_{\nu_{\tau}}}$ & $10^{13}$ GeV\\
\hline
$g_{app}$ & -0.440\\
\hline
$g_{ann}$ & 0.58\\
\hline
$E$ & $-1/3$\\
\hline
$g_{a\gamma}$ & $-3\times 10^{-14}$ GeV$^{-1}$\\
\hline
\end{tabular}
\caption{Model B}\label{T:modelB}
\end{table}

Notice that as seen in table \ref{T:modelB}, the neutrino masses arise from dimension 5 operators. Thus they are only suppressed by the ratio $\langle h\rangle/M$ relative to the top quark, and this pushes the scale $M$ higher. Also, all leptons have their masses from dimension five operators so they are naturally light. The string scale is compatible with expectations from D-brane intermediate scale models. At this stage we can declare success for this style of model building.

\subsection{Three-Stack Models}

\subsubsection{Model C}

This model is problematic: the FN terms split the quark isospin doublets between the light and heavy tiers, with $\frac{m_{b}}{m_{t}}$ controlling the difference between the mass tiers and fails to explain why some of the quarks are light by much larger fine tunings than in our previous examples. Just based on this one should strongly disfavor  models like this one.
Despite the schizophrenic pattern of Yukawa couplings, which is no better than those of the standard model, model C does however have the advantage of suppressing lepton masses and sharing important characteristics with model $B$.

The anomaly-free hypercharge is given by $Q_{Y}=\frac{1}{6}(Q_{c}-3Q_{b})$. The orthogonal anomalous charges are

\[
\tilde{Q}_{1}=\frac{8+7\sqrt{2}}{6}Q_{w}-\frac{1}{3\sqrt{2}}Q_{b}+Q_{c},
\qquad \tilde{Q}_{2}=\frac{8-7\sqrt{2}}{6}Q_{w}+\frac{1}{3\sqrt{2}}Q_{b}+Q_{c}
\]
We choose $Q_{PQ}=\frac{1}{2}(Q_{w}+Q_{b})$ in this model. Model D has identical anomalous, anomaly-free, and PQ charges.

Once again, table~\ref{T:modelC} indicates that the axion has ordinary couplings to matter, so we conclude that the axion decay constant in model C is close to the upper sill of the axion window, but not greatly outside it, rendering it mostly harmless. Indeed, the upper sill of the axion window depends on typicality arguments of the axion field expectation values in the early universe and should not be used to discard models; it does require additional mechanisms in the early universe that prevent the axion from acquiring random expectation values when the axion field decouples from the rest of matter and overclosing the universe.

\begin{table}
\centering
\begin{tabular}{|c|c|}
\hline
Dimension 4 couplings & $h Q \bar{u} + h^{\dagger} q_{I} \bar{d}$\\
\hline
High dimension couplings & $\phi h q_{I}\bar{u} + \phi h^{\dagger} Q\bar{d} + \phi h^{\dagger}l\bar{e} + (hl)^{2}$\\
\hline
$\vev{\phi}\simeq\frac{m_{b}}{m_{t}}\frac{m_{t}^{2}}{m_{\nu_{\tau}}}$ & $10^{12}$ GeV \\
\hline
$M\simeq\frac{m_{t}^{2}}{m_{\nu_{\tau}}}$ & $10^{13}$ GeV\\
\hline
$g_{app}$ & -0.192\\
\hline
$g_{ann}$ & 0.187\\
\hline
$E$ & $55/24$\\
\hline
$g_{a\gamma}$ & $3\times 10^{-15}$ GeV$^{-1}$\\
\hline
\end{tabular}
\caption{Model C}\label{T:modelC}
\end{table}


\subsubsection{Model D}

This model has even more problematic mass hierarchies than model C. After all, the lepton masses are not suppressed and we are required to add fine tuning to the model. The same excuses that were used in model $A$ can be argued here.
  Notice that in table~\ref{T:modelD} the axion couplings are identical to those of model C. This is because the splitting of the isospin doublets forces us to identify $Q$ as the third generation and $q_{I}$ as the first and second generation in both models C and D. Even this model renders the FN mechanism essentially useless. Indeed, there are too many fine tunings so the FN mechanism is not really doing what we were set out to do at the beginning. Its axion falls an order of magnitude short of the lower sill of the axion window and one can say that this model is essentially ruled out.

\begin{table}
\centering
\begin{tabular}{|c|c|}
\hline
Dimension 4 couplings & $h q_{I} \bar{u} + h^{\dagger} Q \bar{d} + h^{\dagger}l\bar{e}$\\
\hline
High dimension couplings & $\phi h Q\bar{u}+ \phi h^{\dagger} q_{I}\bar{d} +  (\phi hl)^{2}$\\
\hline
$\vev{\phi}\simeq\frac{m_{b}}{m_{t}}\frac{m_{b}^{2}}{m_{\nu_{\tau}}}$ & $10^{8}$ GeV \\
\hline
$M\simeq\frac{m_{b}^{2}}{m_{\nu_{\tau}}}$ & $10^{9}$ GeV\\
\hline
$g_{app}$ & -0.192\\
\hline
$g_{ann}$ & 0.187\\
\hline
$E$ & $55/24$\\
\hline
$g_{a\gamma}$ & $3\times 10^{-11}$ GeV$^{-1}$\\
\hline
\end{tabular}
\caption{Model D}\label{T:modelD}
\end{table}

The upshot of this analysis is that for each model the axion couplings are $\mathcal{O}(1)$, and there is no accidental fine-tuning which might radically alter the lower sill of the axion window for each model. Thus, we find that these viable models result in one harmful, one harmless, and two mostly harmless axions.


\section{General considerations and Heavy Axions}

So far, we have built four simple models of flavor textures by following the D-brane effective field theory paradigm. General models will be more complicated, but there are universal features worth discussing. First, considering the general procedure we have used so far, the CKM matrix is completely undetermined. This is either because the left handed quark doublets all appear on the same footing in the Lagrangian (the four stack models), or because the right handed fermions all appear on the same footing in the Lagrangian (the three stack models). It would be nice if the explanation for the CKM matrix was part of the phenomenology addressed by the Froggatt-Nielsen mechanism itself and would not require an independent solution.  In many top-down string theory models this is done geometrically, the various Yukawa couplings are related to areas of triangles  (see \cite{Cvetic,Honecker} for recent progress in the computation of Yukawa couplings in these models), but geometry by itself usually goes beyond effective field theory. Some of these textures might also arise from discrete symmetries, but we are trying to avoid that kind of argument: to our knowledge there is no satisfactory solution of the CKM matrix at this level. Indeed, the Frogatt-Nielsen mechanism is the preferred way to address this issue in phenomenology anyhow.

The second statement that is important to understand is that by using the three and four stack models, independently of how the scalar sector is realized in perturbative D-brane models, we are committing ourselves to a low energy effective field theory with Peccei-Quinn symmetries that are only broken non-perturbatively. These approximate global symmetries are always generated by the Iba\~nez-Quevedo mechanism. Thus the low energy effective field theory will necessarily have light axions arising from the phase factor of the order parameter that breaks the Froggatt-Nielsen symmetry: this can be a complicated linear combination of the scalars that are present, and might even belong to the Higgs sector of the low energy effective field theory. Many of these models are ruled out by current data and therefore axion constraints should be part of the analysis of top down approaches to understand the Standard model or its supersymmetric versions. The Iba\~nez-Quevedo mechanism works just as well in supersymmetric models as in non-supersymmetric models and is a property of the perturbative structure of open string models.

Also notice that as far as low energy effective field theories are concerned, the important detail to determine the axion model is determined by the axion photon couplings $g_{a\gamma}$ and the axion decay constant. Together they determine the mass of the axion. The most important relation for comparison to experiments like CAST is given by
\begin{equation}
g_{\alpha\gamma} \simeq \frac{\alpha}{2\pi} \left( \frac E N - 1.92\right) \frac{m_a}{m_\pi f_\pi}
\end{equation}
For Grand Unified models $E/N\simeq 8/3$ (see for example \cite{Srednicki_85}) producing a slight cancellation in the axion photon coupling , while for the KSVZ model \cite{Kim_79,Shifman_79} $E/N=0$. In our cases $E/N-1.92$ is larger than the KSVZ model, and can be twice  as large. This makes these models more  sensitive to the constraints from experiments like CAST \cite{CAST}, although our considerations put the axion photon coupling for reasonable models in the range $10^{-15} $--$10^{-11}$, but the precise value of the decay constant depends on values of coupling constants that we have set to be of order one, but as yet unspecified. If moreover axions comprise the dark matter halo, then our numbers can be compared with the results of ADMX \cite{Asztalos:2009yp} and the parameter space for our models can get constrained.

A second point that is important to consider is that we assumed at the beginning that the Frogatt-Nielsen  symmetry was being gauged was anomalous, and the anomaly was cancelled with an axion-like field that got eaten up to make a massive $Z'$ near the string scale. This is a closed string field.  However, the mass of the $Z'$ receives two contributions, one from the Green-Schwarz anomaly cancellation mechanism, and another from the usual Higgs mechanism from breaking the FN-symmetry. If the mass of the $Z'$ is dominated by the Higgs term, the low energy axion that survives is the one arising from the closed string sector. The associated axion decay constant is then bounded above by the mass of the $Z'$. This means that the closed string axion would have a relatively low decay constant and might fall outside the axion window. Thus in the general theory analysis it is possible to push down the axion decay constant, but this does not push it up. Indeed, the mass of the $Z'$ fields arises from the terms
\begin{equation}
\frac 12 K_{ij} ( \tilde Z_\mu^i)( \tilde Z_\mu^j)+  \vev{\phi}^2 (\tilde Z^{FN}_\mu)^2
\end{equation}
In the case the second terms dominates, the quantity $K$ is bounded above by $\vev{\phi}$, and $K$ plays the role of the axion decay constant for the closed string modes.
Obviously the mixings complicate the analysis substantially and one also expects that the effective field theory might  break down before the Frogatt-Nielsen scale is reached, making it necessary to build a more complete string model. This is outside the effective field theory paradigm we have been pursuing, but it is a distinct possibility. Fortunately one can not hide the axion this way so the models are testable.

The axions appearing in these models can trace their origins to stringy effects, so it is useful to consider other stringy contributions to the axion mass. Worldsheet, D-brane, and other instantons can have possibly large, nonperturbative contributions to the axion potential. Combining the effects of stringy instantons and QCD, axions develop a potential of the form~\cite{Svrcek_Witten_06}

\[
\widetilde{V}(a) \sim -2 M_{s}^{4}e^{-S_{inst}}\cos(a/f_{a}+\psi)
+ a^{2}\frac{m_{\pi}^{2}f_{\pi}^{2}}{f_{a}^{2}}
\]

All quantities in the potential are predetermined in the models discussed here, so we can treat $S_{inst}$ as a free parameter. Ordinarily, the instanton action is bounded below in order for the axion to solve strong CP. This bound is\footnote{If $M_{s}$ is taken to be the Planck scale, $S_{inst} \gtrsim 200$.}

\[
S_{inst} \gtrsim \log\left(\frac{M_{s}^{4}}{10^{-11}m_{\pi}^{2}F_{\pi}^{2}}\right)
\sim
\left\{\begin{array}{cc}
60 & (\textrm{model A})\\
125 & (\textrm{model B})
\end{array}\right.
\]

In the context of the models discussed here, however, one may take the view that axions are simply unwanted byproducts of the model, and need not solve the strong CP problem. There is then freedom to consider cases where stringy axion mass contributions are important. If the stringy effects dominate, this divorces the relationship between mass and decay constant that is characteristic of the QCD axion (although the couplings all still depend on $f_{a}$). If the axion can be made heavy enough, $m_{a}\geq 100$ MeV, this can avoid the usual astrophysical constraints, and save e.g. model A (models C and D should still be disfavored because of their poorly-behaved mass hierarchies). This gives an upper bound on the instanton action

\[
S_{inst} \lesssim -2\log\left(\frac{10^{-1}\textrm{GeV}f_{a}}{M_{s}^{2}}\right)
\sim
\left\{\begin{array}{cc}
45 & (A)\\
75 & (B)
\end{array}\right.
\]

For the models at hand the bounds on $S_{inst}$ for QCD and heavy axions are incompatible, so these axions cannot simultaneously resemble QCD axions while avoiding the astrophysical constraints.

However, the decays of heavy axions run the risk of interfering with big bang nucleosynthesis. To avoid this, one can require that primordial axions decay away before the nucleosynthesis era $t_{N} \simeq 10^{1}-10^{2}$ s, i.e. $\Gamma_{a\rightarrow X} t_{N} \gg 1$. Photonic decays dominate, since the branching ratio to fermionic decays is

\[
\frac{\Gamma_{a\overline{f}f}}{\Gamma_{a\gamma\gamma}} = \frac{X_{A}^{2}}{4g_{a\gamma\gamma}^{2}}
\left(\frac{m_{f}}{m_{a}}\right)^{4}
\sqrt{1-\left(\frac{2m_{f}}{m_{a}}\right)^{2}}
\]

Since $X_{A}$ is of the same order as $g_{a\gamma\gamma}$, this branching ratio is very small, peaking at $\sim 10^{-3}$. Writing $M_{s}\sim 10^{m}$ GeV, $f_{a}\sim10^{f}$ GeV, and $\hbar \sim 10^{-24}$ GeVs we have

\[
\frac{\Gamma_{a\rightarrow\gamma\gamma}t_{N}}{\hbar} \sim
\frac{m_{a}^{3}}{f_{a}^{2}} \frac{t_{N}}{\hbar} \sim
 e^{-3S_{inst}/2} \frac{M_{s}^{6}}{f_{a}^{5}} \frac{t_{N}}{\hbar} \sim
  10^{6m-5f} 10^{2 + 24} e^{-3S_{inst}/2} \\
\]

This places the constraint on the instanton action

\[
S \ll \frac{2}{3}(26 + 6m -5f)\log 10 \sim
\left\{\begin{array}{cc}
70 & (A)\\
75 & (B)
\end{array}\right.
\]

For model B, this bound matches the bound necessary to have heavy axions in the first place, while for model A the astrophysical bound on $S_{inst}$ is more stringent, but it is certainly possible for these heavy axions to decay fast enough in both cases. If we think of these potentials as generated from an anomaly, we would need a hidden sector for which the $U(1)_{PQ}$ symmetry is anomalous, but the corresponding fermions are arranged into Dirac representations of the standard model. Such a model is not minimal, but it can serve as a field theory realization of this idea that does not require strictly stringy instantons.

Another possibility is that heavy axions are very long-lived, and comprise dark matter. Sikivie \cite{Sikivie_10,Sikivie_Yang_09} has argued that caustics in the dark matter halo are indicative of a Bose-Einstein condensate of axion-like particles. If so, then we must have $\Gamma t_{now} \ll 1$. Unfortunately for heavy axions, their large mass makes the mean lifetime is rather short. For axions with $m_{a}=100$ MeV,

\[
\frac{\hbar}{\Gamma} \sim \frac{\hbar f_{a}^{2}}{m_{a}^{3}} \sim
\left\{\begin{array}{cc}
10^{-15}\;{\rm s} & (A)\\
15\;{\rm s} & (B)
\end{array}\right.
\]

This problem only gets worse the heavier the axion. Trying to avoid the astrophysical constraint on $f_{a}$ by allowing for heavy axions can not result in heavy axion dark matter in the models considered here, since $f_{a}$ is not a free parameter.

Model A is ordinarily ruled out if the axion is a QCD axion, but if its mass depends more strongly on stringy instanton effects than its decay constant, the lower bound on $f_{a}$ can be relaxed. This is allowable for fast-decaying axions, which means that the action of the instantons contributing to the axion potential must be rather small, in stark contrast to the QCD axion. Nevertheless, it is thus possible to open the axion window wide, and allow model A through.

A second problem that we have alluded to before is that the CKM matrix is not part of our proposal for addressing the flavor problem. We would like to comment now on how to remedy this situation. The sticky point in our models is that either the left or the right handed quarks are essentially indistinguishable from each other as far as their quantum numbers are concerned. Thus a model that does better needs to distinguish both the left handed quarks and the right handed quarks. Let us assume that we want to do this in some minimal way. What we need is some gauge theoretic reason to connect some left handed quarks preferentially to some right handed quarks and not the others.

Again, we want to achieve this by using a gauge symmetry reason. Since the quarks are attached to the same $U(3)$ node of QCD, this is the place where we can make the biggest difference, thus we would want an extension of the $U(3)$ color to a bigger gauge group. This should be chiral, and therefore should be represented by products of
$U(n_i)$ gauge groups. These typically break to diagonal subgroups with standard embeddings due to the D-brane matter content.
The simplest scenario is to have a $U(3)\times U(3)$ symmetry where the top and bottom quark are attached to one $U(3)$ and the other quarks are attached to the other $U(3)$. We also need a scalar that breaks to the diagonal $U(3)\times U(3)\to U(3)$. The natural scalar is in the $(3, \bar 3)$ representation. Notice that this introduces a new $U(1)$ gauge group that might suffer anomalies as well and similar reasonings as before (using the Iba\~nez-Quevedo argument) lead to the possibility of a second low energy axion in the dynamics. This extra scalar field would appear in non-renormalizable terms in the lagrangian that provide for the smallness of the mixings between the third family and the other two families. This is indeed what a deconstruction of an extra dimension for QCD would look like \cite{ArkaniHamed:2001ca} (supersymmetric extensions of the standard model that have this flavor structure have been considered recently in \cite{Craig:2011yk} ).  This is only one step away from thinking about this problem using large extra dimensions as they appear in a full compactification of string theory and suggests that in the end this problem can always  be argued to be resolved geometrically.
 Naturally, the analysis here becomes more complicated and will not be pursued further in this paper.

\section{Conclusion}

We have considered using the Froggatt-Nielsen mechanism to generate the minimal amount of hierarchy between fermion masses, separating them into light and heavy tiers. The novelty of our construction is that the Frogatt-Nielsen symmetry has mixed anomalies that are cancelled with the Green-Schwarz mechanism. This anomaly cancellation provides a Stuckelberg mass for the $Z'$ associated to this symmetry. The peculiarities of  realizations of this class of models in perturbative D-brane constructions leave a residue of the Froggatt-Nielsen symmetry at low energies that acts as a Peccei-Quinn symmetry. When the symmetry is broken in order to account for flavor hierarchies, a pseudo-goldstone boson is generated which behaves like an axion. This arises from the open string sector dynamics and can be considered as an open string axion. We do an effective field theory analysis of the physics by considering non-renormalizable contributions to the lagrangian suppressed by a high energy mass scale which we take to be the string scale. The ratio of the Frogatt-Nielsen order parameter to the string scale is determined by the quark mass ratios.
The Froggatt-Nielsen fields can also participate in how the neutrino masses are generated, so it is natural to consider the neutrino masses to calculate the high scale. If the neutrino masses are generated by dimension five operators, the high energy string scale is of order $10^{13}$ GeV, as is typical of large extra dimension models based on D-branes. If the neutrino mass is generated by dimension $7$ operators the string scale is much lower and depends on details of the Froggatt-Nielsen suppressions of a given model. In general, the Froggatt-Nielsen order parameter serves as a benchmark for the axion decay constant and it is between one and three orders of magnitude below the string scale. Many possible realizations of the standard model within string theory could be ruled out this way, as they would lead to unacceptably small axion decay constants outside the allowed axion window. This issue makes it clear that axion bounds need to be analyzed in order to decide if D-brane models built from string theory are truly viable.
Furthermore, our analysis predicts relatively large axion-photon-photon couplings so the models can be compared to experimental searches like those carried at the CAST and ADMX experiments. This gives us a prospect for ruling out such extensions of the standard model.

There are ways out of this dilemma that would require additional stringy instantons to make the low energy axions heavier in order to avoid astrophysical constraints from nuclear physics processes.
We also found that in the simplest models we consider we have nothing to say about the CKM matrix and to find an acceptable solution that includes this information it is likely that one needs to deconstruct extra dimensions so that strong QCD dynamics arises from a $U(3)\times U(3)\to U(3)$ gauge symmetry breaking at high energies or an even bigger extension. The details of such models are beyond the scope of the present paper.

If one is willing to give up the possibility of a string embedding, one can assign arbitrary $U(1)_{a,b}$ charges to the various fields; this allows higher-dimensional FN terms of the generic forms $\phi^{q}\left(hq_{L}q_{R}\right)$ and $\phi^{n}(h l_{L})^{2}$, giving masses $m_{q}\simeq \langle h\rangle\left(\frac{\vev{\phi}}{M}\right)^{q},\,m_{\nu}\simeq \frac{\langle h\rangle^{2}}{M}\left(\frac{\vev{\phi}}{M}\right)^{n}$ for light quarks and neutrinos, respectively. Both $M$ and $\vev{\phi}$ can be estimated from combining the information of these ratios. In particular, the most important for us is the axion decay constant which is essentially given by the vacuum expectation value $\langle \phi\rangle$. This is given by
\[
\vev{\phi} \simeq m_{t}^{(2q-n-1)/q}m_{q}^{(n+1)/q}m_{\nu_{\tau}}^{-1}
\]
where $m_{q}$ is the mass of the heaviest light quark. This is very model dependent. In general one obtains a lot of possibilities this way and one needs to pick the one that corresponds to the highest value of $\vev{\phi}$, by assuming that the other couplings that give rise to a high scale have extra fine tunings that make them smaller
than our naive guess would suggest.

To conclude, we find that some string inspired minimal models for flavor are able to predict axions within the allowed axion window by assuming that the scale of new physics (taken as the string scale) is closely related to the scale at which the neutrino masses are generated.
The models also predict an intermediate string scale that can be rather light in the  neutrino masses arise from high dimension operators (dimension 7). If the string scale is too light, the axion falls outside the allowed window. In each case, the models fail to explain the smallness of some of the Yukawa couplings. It would be interesting to study further how such models can be embedded in supersymmetric setups so that the hierarchy problem can also be addressed.

\pp
{\em Acknowledgements:} We would like to thank M. Srednicki and M. Cvetic for discussions. Work supported in part by DOE under grant DE-FG02-91ER40618. D.B. work supported in part by the National Science Foundation under Grant No. NSF Phy05-51164



\begin{thebibliography}{99}

\bibitem{Banks:2010zn}
  T.~Banks and N.~Seiberg,
 ``Symmetries and Strings in Field Theory and Gravity,''
  Phys.\ Rev.\ D {\bf 83}, 084019 (2011)
  [arXiv:1011.5120 [hep-th]].

\bibitem{Ibanez_Quevedo_99}
  L.~E.~Ibanez and F.~Quevedo,
  ``Anomalous U(1)'s and proton stability in brane models,''
  JHEP {\bf 9910}, 001 (1999)
  [arXiv:hep-ph/9908305].


\bibitem{Froggatt_Nielsen_79}
  C.~D.~Froggatt and H.~B.~Nielsen,
  ``Hierarchy Of Quark Masses, Cabibbo Angles And CP Violation,''
  Nucl.\ Phys.\  B {\bf 147}, 277 (1979).

\bibitem{Berenstein_Perkins_10}
D.~Berenstein and E.~Perkins,
  ``A viable axion from gauged flavor symmetries,''
  Phys.\ Rev.\ D {\bf 82}, 107701 (2010)
  [arXiv:1003.4233 [hep-th]].


\bibitem{Green_Schwarz_84}
  M.~B.~Green and J.~H.~Schwarz,
  ``Anomaly Cancellation In Supersymmetric D=10 Gauge Theory And Superstring
  Theory,''
  Phys.\ Lett.\  B {\bf 149}, 117 (1984).



\bibitem{Peccei_Quinn_77}
  R.~D.~Peccei and H.~R.~Quinn,
  ``CP Conservation In The Presence Of Instantons,''
  Phys.\ Rev.\ Lett.\  {\bf 38}, 1440 (1977).

\bibitem{Weinberg_78}
  S.~Weinberg,
  ``A New Light Boson?,''
  Phys.\ Rev.\ Lett.\  {\bf 40}, 223 (1978).

\bibitem{Wilczek_78}
  F.~Wilczek,
  ``Problem Of Strong P And T Invariance In The Presence Of Instantons,''
  Phys.\ Rev.\ Lett.\  {\bf 40}, 279 (1978).

\bibitem{Turner_89}
  M.~S.~Turner,
  ``Windows on the Axion,''
  Phys.\ Rept.\  {\bf 197}, 67 (1990).

\bibitem{Svrcek_Witten_06}
  P.~Svrcek and E.~Witten,
  ``Axions in string theory,''
  JHEP {\bf 0606}, 051 (2006)
  [arXiv:hep-th/0605206].


\bibitem{Kiritsis1} 
  I.~Antoniadis, E.~Kiritsis, J.~Rizos and T.~N.~Tomaras,
  ``D-branes and the standard model,''
  Nucl.\ Phys.\ B {\bf 660}, 81 (2003)
  [hep-th/0210263].


\bibitem{Kiritsis2} 
  C.~Coriano, N.~Irges and E.~Kiritsis,
  ``On the effective theory of low scale orientifold string vacua,''
  Nucl.\ Phys.\ B {\bf 746}, 77 (2006)
  [hep-ph/0510332].

\bibitem{Blumenhagen_et_al_05}
  R.~Blumenhagen, M.~Cvetic, P.~Langacker and G.~Shiu,
  ``Toward realistic intersecting D-brane models,''
  Ann.\ Rev.\ Nucl.\ Part.\ Sci.\  {\bf 55}, 71 (2005)
  [arXiv:hep-th/0502005].

\bibitem{Berenstein_Pinansky_07}
  D.~Berenstein and S.~Pinansky,
  ``The Minimal Quiver Standard Model,''
  Phys.\ Rev.\  D {\bf 75}, 095009 (2007)
  [arXiv:hep-th/0610104].

\bibitem{Coriano_Irges_06}
  C.~Coriano and N.~Irges,
  ``Folding Froggatt-Nielsen into the Stueckelberg-Higgs mechanism in anomalous
  U(1) models,''
  arXiv:hep-ph/0612128.

\bibitem{Blumenhagen_et_al_09}
  R.~Blumenhagen, M.~Cvetic, S.~Kachru and T.~Weigand,
  ``{\small D}-Brane Instantons in Type {II} Orientifolds,''
  Ann.\ Rev.\ Nucl.\ Part.\ Sci.\  {\bf 59}, 269 (2009)
  [arXiv:0902.3251 [hep-th]].

\bibitem{Berezhiani_83}
  Z.~G.~Berezhiani,
  ``The weak mixing angles in gauge models with horizontal symmetry - a new approach to quark and lepton masses,''
  Phys. \ Lett.\ B {\bf 129}, 99 (1983)

\bibitem{Berezhiani_Khlopov_91}
  Z.~G.~Berezhiani and M.~Yu.~Khlopov,
  ``Cosmology of spontaneously broken gauge family symmetry with axion solution of strong CP problem,''
 Z. \ Phys.\ C {\bf 49}, 73 (1991)

\bibitem{Berenstein:2006aj}
  D.~Berenstein,
  ``Branes versus GUTS: Challenges for string inspired phenomenology,''
  hep-th/0603103.

\bibitem{Kiritsis3} 
  P.~Anastasopoulos, E.~Kiritsis and A.~Lionetto,
  ``On mass hierarchies in orientifold vacua,''
  JHEP {\bf 0908}, 026 (2009)
  [arXiv:0905.3044 [hep-th]].

\bibitem{Cvetic2} 
  M.~Cvetic, J.~Halverson and R.~Richter,
  ``Mass Hierarchies versus proton Decay in MSSM Orientifold Compactifications,''
  arXiv:0910.2239 [hep-th].

\bibitem{Antoniadis_et_al_00}
  I.~Antoniadis, E.~Kiritsis and T.~N.~Tomaras,,
  ``A D-brane alternative to unification,''
  Phys.\ Lett.\ B {\bf 486}, 186 (2000)
  [arXiv:0004214 [hep-ph]].

\bibitem{Kim_79}
  J.~E.~Kim,
  ``Weak Interaction Singlet And Strong CP Invariance,''
  Phys.\ Rev.\ Lett.\  {\bf 43}, 103 (1979).

\bibitem{Shifman_79}
  M.~A.~Shifman, A.~I.~Vainshtein and V.~I.~Zakharov,
  ``Can Confinement Ensure Natural CP Invariance Of Strong Interactions?,''
  Nucl.\ Phys.\  B {\bf 166}, 493 (1980).

\bibitem{Zhitnitsky_80}
  A.~R.~Zhitnitsky,
  ``On Possible Suppression Of The Axion Hadron Interactions. (In Russian),''
  Sov.\ J.\ Nucl.\ Phys.\  {\bf 31}, 260 (1980)
  [Yad.\ Fiz.\  {\bf 31}, 497 (1980)].

\bibitem{Dine_et_al_81}
  M.~Dine, W.~Fischler and M.~Srednicki,
  ``A Simple Solution To The Strong CP Problem With A Harmless Axion,''
  Phys.\ Lett.\  B {\bf 104}, 199 (1981).

\bibitem{Srednicki_85}
  M.~Srednicki,
  ``Axion Couplings To Matter. 1. CP Conserving Parts,''
  Nucl.\ Phys.\  B {\bf 260}, 689 (1985).

\bibitem{Georgi_et_al_86}
  H.~Georgi, D.~B.~Kaplan and L.~Randall,
  ``Manifesting The Invisible Axion At Low-Energies,''
  Phys.\ Lett.\  B {\bf 169}, 73 (1986).

\bibitem{Kaplan_85}
  D.~B.~Kaplan,
  ``Opening The Axion Window,''
  Nucl.\ Phys.\  B {\bf 260}, 215 (1985).

\bibitem{PDG}
  C.~Amsler {\it et al.}  [Particle Data Group],
  ``Review of particle physics,''
  Phys.\ Lett.\  B {\bf 667}, 1 (2008).

\bibitem{Kim_C_08}
  J.~E.~Kim and G.~Carosi,
  ``Axions and the Strong CP Problem,''
  arXiv:0807.3125 [hep-ph].

\bibitem{Brinkmann_Turner_88}
  R.~Brinkmann and M.~Turner,
  ``Numerical rates for nucleon-nucleon, axion bremsstrahlung,''
  Phys.\ Rev.\  D {\bf 38}, 2338 (1988)

\bibitem{Dashen_69}
  R.~Dashen,
  ``Chiral SU(3)xSU(3) as a symmetry of the strong interactions,''
  Phys.\ Rev.\  {\bf 183}, 1245 (1969)


\bibitem{CAST}
  E.~Arik {\it et al.}  [CAST Collaboration],
  ``Probing eV-scale axions with CAST,''
  JCAP {\bf 0902}, 008 (2009)
  [arXiv:0810.4482 [hep-ex]].
  F.~J.~Iguaz [CAST Collaboration],
  ``The CAST experiment: Status and perspectives,''
  PoS IDM {\bf 2010}, 056 (2011)
  [arXiv:1110.2116 [hep-ex]].

\bibitem{Asztalos:2009yp} 
  S.~J.~Asztalos {\it et al.}  [The ADMX Collaboration],
  ``A SQUID-based microwave cavity search for dark-matter axions,''
  Phys.\ Rev.\ Lett.\  {\bf 104}, 041301 (2010)
  [arXiv:0910.5914 [astro-ph.CO]].

\bibitem{Sikivie_10}
  P. Sikivie,
  ``The emerging case for axion dark matter,''
  Phys.\ Lett.\  B {\bf 667}, 1 (2008)..

\bibitem{Sikivie_Yang_09}
  P.~Sikivie and Q.~Yang,
  ``Bose-Einstein condensation of dark matter axions,''
  arXiv:1003.2426 [astro-ph.GA].

\bibitem{Cvetic} 
  M.~Cvetic, J.~Halverson and R.~Richter, 2,
  ``Realistic Yukawa structures from orientifold compactifications,''
  JHEP {\bf 0912}, 063 (2009)
  [arXiv:0905.3379 [hep-th]].


\bibitem{Honecker} 
  G.~Honecker and J.~Vanhoof,
  ``Yukawa couplings and masses of non-chiral states for the Standard Model on D6-branes on T6/Z6',''
  arXiv:1201.3604 [hep-th].
  G.~Honecker and J.~Vanhoof,
  ``Towards the field theory of the Standard Model on fractional D6-branes on T6/Z6': Yukawa couplings and masses,''
  arXiv:1201.5872 [hep-th].
  
\bibitem{ArkaniHamed:2001ca}
  N.~Arkani-Hamed, A.~G.~Cohen and H.~Georgi,
  ``(De)constructing dimensions,''
  Phys.\ Rev.\ Lett.\  {\bf 86}, 4757 (2001)
  [hep-th/0104005].

\bibitem{Craig:2011yk} 
  N.~Craig, D.~Green and A.~Katz,
  ``(De)Constructing a Natural and Flavorful Supersymmetric Standard Model,''
  JHEP {\bf 1107}, 045 (2011)
  [arXiv:1103.3708 [hep-ph]].



\end{thebibliography}
\end{document}